\documentclass[twocolumn,showpacs,prd,floatfix,axodraw]{revtex4}
\usepackage{mathrsfs}
\usepackage{graphicx,,booktabs,bm}
\usepackage{overpic}
\usepackage{color}
\usepackage{amssymb}

\usepackage{mathrsfs,bm,amsmath,amssymb}
\usepackage{longtable,lscape}
\usepackage{txfonts}
\usepackage{amssymb}
\usepackage{indentfirst}
\usepackage{graphicx,,booktabs}
\usepackage{multirow,ulem}

\usepackage{color}
\usepackage{amssymb}

\definecolor{cover}{rgb}{0.77,0.87,0.88}
\definecolor{blueone}{rgb}{0.1,0.1,.7}
\definecolor{citec}{rgb}{0.14,0.47,0.09}
\definecolor{two}{rgb}{0.0,0.5,0.}
\definecolor{three}{rgb}{.5,.1,0.15}
\usepackage[bookmarks=true,bookmarksopen=false,plainpages=false,breaklinks=true,
   bookmarksnumbered=true,hypertexnames=false,
   filecolor=blue,urlcolor=three,menucolor=three,
   linkcolor=three,citecolor=blueone, colorlinks,
   anchorcolor=blue,runcolor=pink,frenchlinks=red
   pdfstartview=FitH,pdftitle=title,%
   pdfauthor=author]{hyperref}

\begin{document}
\title{Strong decays of the  $\Xi_b(6227)$ as a $\Sigma_b\bar{K}$ molecule }

\author{Yin Huang}
\affiliation{School of Physics and Nuclear Energy Engineering, Beihang University, Beijing 100191, China}

\author{Cheng-jian Xiao}
\affiliation{Institute of High Energy Physics, Chinese Academy of Sciences, Beijing 100049, China}

\author{Li-Sheng Geng\footnote{corresponding author}} \email{lisheng.geng@buaa.edu.cn}
\affiliation{School of Physics and Nuclear Energy Engineering,
International Research Center for Nuclei and Particles in the Cosmos
and Beijing Key Laboratory of Advanced Nuclear Materials and
Physics, Beihang University, Beijing 100191, China}

\author{Jun He}\email{junhe@njnu.edu.cn}
\affiliation{Department of Physics and Institute of Theoretical Physics, Nanjing Normal University,
Nanjing 210097, People¡¯s Republic of China}

\begin{abstract}
We study the strong decays of the newly observed $\Xi_b(6227)$ assuming  that it is a pure $\Sigma_b\bar{K}$  molecular state.
Considering four possible spin-parity assingments $J^P=1/2^{\pm}$ and $3/2^{\pm}$, the partial decay widths of the $\Sigma_b\bar{K}$ molecular state into  $\Lambda_b\bar{K}$, $\Xi_b\pi$, and $\Xi_b^{'}\pi$ final states through hadronic loops are evaluated with the help of the effective Lagrangians.
In comparison with the LHCb data, the spin-party $1/2^-$ assignment is preferred while those of  $J^P=1/2^{+}$ and $J^P=3/2^{\pm}$ are disfavored.  In addition, we show that
the two allowed decay modes $\Lambda_b\bar{K}$ and $\Xi_b\pi$ of the $\Xi_b(6227)$, being a $S$-wave $\Sigma_b\bar{K}$ molecular state,  have almost equal branching ratios,
consistent with the data.\end{abstract}

\date{\today}


\maketitle
\section{Introduction}
During the past decade, many new narrow baryon resonances  containing a $b$ quark, such as $\Lambda_b(5912)$, $\Lambda_b(5920)$, $\Sigma_b(5815)$, $\Sigma^{*}_b$, and $\Xi_b$ , were discovered
experimentally~\cite{Aaij:2012da,Aaltonen:2007ar,Aaij:2014lxa}.   Very recently, a bottom baryon $\Xi_b(6227)$ was reported by the LHCb Collaboration in both the $\Lambda_b^{0}K^{-}$ and $\Xi^{0}_b\pi^{-}$ final states from $pp$ collisions~\cite{Aaij:2018yqz}.  Its mass and width are, respectively,
\begin{align}
&M=6226.9\pm{2.0}(stat)\pm0.3(syst)\pm{}0.2(\Lambda^0_b)~ {\rm MeV}\nonumber\\
&\Gamma=18.1\pm5.4(stat)\pm1.8(syst)~{\rm MeV}.
\end{align}
Its spin-parity, however,  remains unknown.

Before the discovery of the $\Xi_b^{*}$ state,  there were already  a few theoretical studies on the existence of such a state.  For instance, different  quark models predicted that there exists a $dsb$ three quark excited state with a mass about 6200 MeV~\cite{Ebert:2007nw,Ebert:2011kk,Thakkar:2016dna}.
The mass spectra of bottom baryons around 6200 MeV were investigated in a Faddeev equation approach~\cite{Valcarce:2008dr}.
A molecule $\Xi_b$ state with a narrow width and a mass around 6200 MeV was predicted in the unitarized coupled channels approach~\cite{GarciaRecio:2012db}.
In Ref~\cite{Lu:2014ina} the authors also found that the $\bar{K} \Sigma_b$ interaction  is strong enough to  form a bound state  both in isospin $3/2$ and $1/2$.~\footnote{We note that
in Ref.~\cite{Lu:2014ina},  the predicted $\bar{K}\Sigma_b$ molecular state in isospin $1/2$ is much lower than that in isospin $3/2$, whose position is closer to that of the
$\Xi_b^*$. }

Following the discovery of the $\Xi_b^{*}$, several theoretical studies  have been performed~\cite{Wang:2018fjm,Chen:2018orb,Aliev:2018lcs}. In the heavy quark-light diquark
model, based on the analysis of the mass spectrum and the two-body OZI-allowed strong decays, the $\Xi_b^{*}$ was suggested to be a $1P$ $\Xi_b^{'}$ state with spin-parity $J^P = 3/2^-$ or $J^P = 5/2^-$~\cite{Chen:2018orb}.
In the chiral quark model, based on the  two-body strong decays studied, Ref.~\cite{Wang:2018fjm} assigned the
$\Xi_b^{*}$ as a three-quark state with spin-parity $J^P = 3/2^-$ or $J^P = 5/2^-$.
While in Ref.~\cite{Aliev:2018lcs} the mass and two-body strong decays of the $\Xi_b^{*}$ were studied  in the QCD sum rules approach and it was shown
that the $\Xi_b(6227)$ might be a $1P$ orbitally excited $\Xi_b$(5955) state with $J^P =3/2^-$.

Although   the studies of Refs.~\cite{Wang:2018fjm,Chen:2018orb,Aliev:2018lcs} seem to indicate that this state is a conventional three-quark state,  the $\Xi_b^{*}$ might still be a $\bar{K}\Sigma_b$ hadronic molecule state, because the mass gap between the
$\Xi_b^{*}$ and the ground $\Xi_b$, about 440 MeV,  is large enough to excite a light quark-antiquark pair to form a molecular state. Indeed,
 it is shown in Refs.~\cite{GarciaRecio:2012db,Lu:2014ina} that the interaction between a $\bar{K}$ meson and a $\Sigma_b$ baryon is strong enough to form a
bound state with a mass about 6200 MeV.   One way to distinguish the two scenarios is to study the two-body strong decay widths of
the $\Xi_b^{*}$ baryon.  In the present paper we consider the following strong decay modes, $\Xi_b^{*}\to\Lambda_b^0\bar{K}$, $\Xi_b^{*}\to\Xi_b\pi$, and $\Xi_b^{*}\to\Xi_b^{'}\pi$,
of the $\Xi_b(6227)$ with the following spin-parity assignments: $J^P=1/2^{\pm}$ and $3/2^{\pm}$,  using an effective Lagrangian approach and assuming that the $\Xi_b(6227)$
is a hadronic molecule state of $\bar{K}$ and $\Sigma_b$.

This work is organized as follows. The theoretical
formalism is explained in Sec. II. The predicted partial
decay widths are presented in Sec. III, followed by a short summary in the last section.

 \section{FORMALISM AND INGREDIENTS}
In this section, we explain how to calculate the strong decay widths $\Xi_b^{*}\to\Lambda_b\bar{K}$, $\Xi_b^{*}\to\Xi_b\pi$, and $\Xi_b^{*}\to\Xi_b^{'}\pi$
in the molecular scenario with different spin-parity assignments for the $\Xi_b^{*}$.   The molecular structure of the $\Xi_b^{*}$ baryon with $J^P=1/2^{\pm}$ is described by the following Lagrangian~\cite{Dong:2010gu}
\begin{align}
{\cal{L}}_{\Xi_b^{*}}(x)&=g_{\Xi_b^{*}\bar{K}\Sigma_b}\int{}d^4y\Phi(y^2)\nonumber\\
                        &\times\bar{K}(x+\omega_{\Sigma_b}y)\Gamma\vec{\tau}\cdot\vec{\Sigma}_b(x-\omega_{K}y)\bar{\Xi}_b^{*}(x)\label{eq1},
\end{align}
while for $J^P=3/2^{\pm}$ the Lagrangian contains a derivative $\bar{K}\Sigma_b$ coupling
\begin{align}
{\cal{L}}_{\Xi_b^{*}}(x)&=g_{\Xi_b^{*}\bar{K}\Sigma_b}\int{}d^4y\Phi(y^2)\nonumber\\
                        &\times\bar{K}(x+\omega_{\Sigma_b}y)\Gamma\vec{\tau}\cdot\partial_{\mu}\vec{\Sigma}_b(x-\omega_{K}y)\bar{\Xi}_b^{*\mu}(x)\label{eq2},
\end{align}
where $\omega_{K}=m_K/(m_K+m_{\Sigma_b})$, $\omega_{\Sigma_b}=m_{\Sigma_b}/(m_K+m_{\Sigma_b})$, and $\Gamma$ is the corresponding Dirac matrix
reflecting the spin-parity of $\Xi_b^{*}$.  In particular, for $J^{p}=1/2^{+}$ and $3/2^{-}$, $\Gamma=\gamma^{5}$, while for $J^{p}=1/2^{-}$ and $3/2^{+}$,
 $\Gamma=1$.  In the Lagrangian, an effective correlation function $\Phi(y^2)$ is introduced to describe the distribution of the two
constituents, the $\bar{K}$ and the $\Sigma_b$, in the hadronic molecular $\Xi_b^{*}$ state.  The introduced correlation function also makes the Feynman diagrams ultraviolet finite. Here we choose the Fourier transformation of the correlation to be a Gaussian form in the Euclidean space~\cite{Faessler:2007gv,Faessler:2007us,Dong:2008gb,Dong:2009uf,Dong:2009yp,Dong:2017rmg,Dong:2014ksa,Dong:2014zka,Dong:2013kta,Dong:2013iqa,Dong:2013rsa,Dong:2012hc,Dong:2011ys,Dong:2010xv,Dong:2010gu,Dong:2009tg,Dong:2017gaw} ,
\begin{align}
\Phi(p_E^2)\doteq\exp(-p_E^2/\Lambda^2)
\end{align}
with $\Lambda$ being the size parameter which characterizes the distribution of the components inside the molecule.
Though the value of $\Lambda$ could not be determined from first principles, it is usually chosen to be about $1$ GeV
in the literature~\cite{Faessler:2007gv,Faessler:2007us,Dong:2008gb,Dong:2009uf,Dong:2009yp,Dong:2017rmg,Dong:2014ksa,Dong:2014zka,Dong:2013kta,Dong:2013iqa,Dong:2013rsa,Dong:2012hc,Dong:2011ys,Dong:2010xv,Dong:2010gu,Dong:2009tg,Dong:2017gaw}. In this work, we vary  $\Lambda$ in a range of 0.9 GeV $\leq\Lambda\leq{}1.10$ GeV.

\begin{figure}[t]
\begin{center}
\includegraphics[scale=0.50]{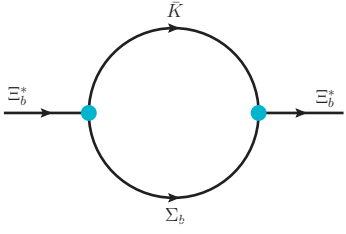}
\caption{Self-energy of the $\Xi_b^{*}$ state.}\label{msef}
\end{center}
\end{figure}

With the effective Lagrangian in Eq.~(\ref{eq1}) and Eq.~(\ref{eq2}) and the Feynman diagram shown in Fig~\ref{msef},
we can obtain the self energy of the $\Xi_b^{*}$,
\begin{align}
\Sigma^{1/2}_{\Xi_b^{*}}(k_0)&=g^2_{\Xi_b^{*}\Sigma_b\bar{K}}\int\frac{d^4k_1}{(2\pi)^4i}\{2\Phi^2[(k_1-k_0\omega_{\Sigma^{-}_b})^2]\Gamma\frac{k\!\!\!/_1+m_{\Sigma_b^{-}}}{k_1^2-m^2_{\Sigma_b^{-}}}\Gamma\nonumber\\
                             &\times\frac{1}{(k_1-k_0)^2-m^2_{\bar{K}^{0}}}+\Phi^2[(k_1-k_0\omega_{\Sigma^{0}_b})^2]\Gamma\frac{k\!\!\!/_1+m_{\Sigma_b^{0}}}{k_1^2-m^2_{\Sigma_b^{0}}}\Gamma\nonumber\\
                             &\times\frac{1}{(k_1-k_0)^2-m^2_{{K}^{-}}}\}\label{eqn1},\\
\Sigma^{\mu\nu3/2}_{\Xi_b^{*}}(k_0)&=g^2_{\Xi_b^{*}\Sigma_b\bar{K}}\int\frac{d^4k_1}{(2\pi)^4i}\{2\Phi^2[(k_1-k_0\omega_{\Sigma^{-}_b})^2]\Gamma\frac{k\!\!\!/_1+m_{\Sigma_b^{-}}}{k_1^2-m^2_{\Sigma_b^{-}}}\Gamma\nonumber\\
                             &\times\frac{1}{(k_1-k_0)^2-m^2_{\bar{K}^{0}}}+\Phi^2[(k_1-k_0\omega_{\Sigma^{0}_b})^2]\Gamma\frac{k\!\!\!/_1+m_{\Sigma_b^{0}}}{k_1^2-m^2_{\Sigma_b^{0}}}\Gamma\nonumber\\
                             &\times\frac{1}{(k_1-k_0)^2-m^2_{{K}^{-}}}\}k_1^{\mu}k_1^{\nu},\label{eqn2}
\end{align}
where $k_0^2=m^2_{\Xi_b^{*}}$ with $k_0, m_{\Xi_b^{*}}$ denoting the four momenta and mass of the $\Xi_b^{*}$, respectively,  $k_1$, $m_{\bar{K}}$, and $m_{\Sigma_b}$ are
the four-momenta, the mass of the $\bar{K}$ meson, and the mass of the $\Sigma_b$ baryon, respectively.  The coupling constant $g_{\Xi_b^{*}\Sigma_b\bar{K}}$ is determined by the compositeness condition~\cite{Dong:2010gu,Faessler:2007gv,Faessler:2007us,Dong:2008gb,Dong:2009uf,Dong:2009yp,Dong:2017rmg,Dong:2014ksa,Dong:2014zka,Dong:2013kta,Dong:2013iqa,Dong:2013rsa,Dong:2012hc,Dong:2011ys,Dong:2010xv,Dong:2009tg,Dong:2017gaw}. It implies that the renormalization
constant of the hadron wave function is set to zero, i.e.,
\begin{align}
Z_{\Xi_b^{*}}=1-\frac{d\Sigma^{1/2(3/2-T)}_{\Xi_b^{*}}}{dk\!\!\!/_0}|_{k\!\!\!/_0=m_{\Xi^{*}_{b}}}=0\label{eqn3},
\end{align}
where the $\Sigma^{3/2-T}_{\Xi_b^{*}}$ is the transverse part of the self-energy operator $\Sigma^{\mu\nu3/2}_{\Xi_b^{*}}$, related to $\Sigma^{\mu\nu3/2}_{\Xi_b^{*}}$ via
\begin{align}
\Sigma^{\mu\nu3/2}_{\Xi_b^{*}}(k_0)=(g_{\mu\nu}-\frac{k_0^{\mu}k_0^{\nu}}{k_0^2})\Sigma^{3/2-T}_{\Xi_b^{*}}+\cdots.\label{eqn4}
\end{align}

The strong decay modes of the $\Xi_b^{*}$ are $\Xi_b^{*}\to\Lambda^{(')0}_bK^{-}$, $\Xi_b^{*}\to\Lambda^{(')0}_b\bar{K}^{0}$,
$\Xi_b^{*}\to\Xi_b^{(')0}\pi^{-}$, $\Xi_b^{*}\to\Xi_b^{(')-}\pi^{0}$,
and $\Xi_b^{*}\to\Xi_b^{(')0}\pi^{0}$.   Here, we only calculate  the
decay widths in the channels $\Xi_b^{*}\to\Lambda_b^{0}K^-$ and $\Xi_b^{*}\to\Xi_b^{(')0}\pi^{-}$.  The decay widths in the other channels can be obtained by isospin symmetry
as $\Gamma(\Xi_b^{*}\to\Lambda_b^{0}\bar{K}^{0})=\Gamma(\Xi_b^{*}\to\Lambda_b^{0}K^-)$ and $\Gamma(\Xi_b^{*}\to\Xi_b^{(')0}\pi^{0})$ = $\Gamma(\Xi_b^{*}\to\Xi_b^{(')-}\pi^{0})= \frac{1}{2}\Gamma(\Xi_b^{*}\to\Xi_b^{(')0}\pi^{-})$.  The sum of these partial decay widths is the total decay width of the $\Xi_b^{*}$.
\begin{figure}[htbp]
\begin{center}
\includegraphics[scale=0.40]{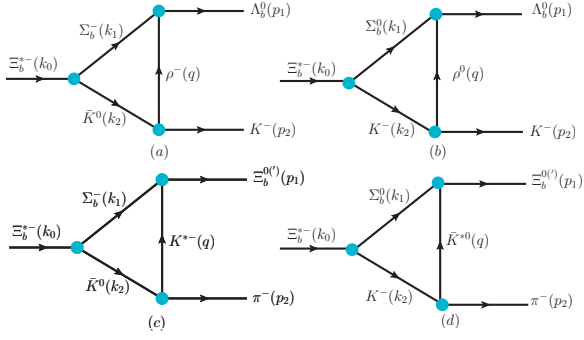}
\caption{Feynman diagrams for the $\Xi_b^{*-}\to{}\Lambda_b^{0}K^{-}$,$\Xi_b^{0}\pi^{-}$, and $\Xi_b^{'0}\pi^{-}$ decay processes.  We also show
the definitions of the kinematics ($k_0,k_1,k_2,p_1,p_2$, and $q$) used in the calculation.}\label{fety}
\end{center}
\end{figure}

Fig.~\ref{fety} shows the hadronic decay of the $\Sigma_b\bar{K}$ molecular state into  $\Lambda_b^0K^{-}$, $\Xi_b^0\pi^{-}$, and $\Xi_b^{0'}\pi^{-}$
mediated by the exchange of the $\rho$ and $\bar{K}^{*}$ mesons.  In Refs.~\cite{Hofmann:2005sw,Montana:2017kjw}, the couplings of the vector meson with
the charm baryons are obtained using the hidden-gauge formalism and assuming SU(4) flavor  symmetry:
\begin{align}
{\cal{L}}_{VBB}=\frac{g}{2}\sum_{i,j,k,l=1}^{4}\bar{B}_{ijk}\gamma^{\mu}(V^{K}_{\mu,l}B^{ijl}+2V^{j}_{\mu,l}B^{ilk}),
\end{align}
where the coupling constant $g=6.6$ is taken from Ref.~\cite{Hofmann:2005sw}.  The symbol $V_{\mu}$ represents the vector fields of the 16-plet of the $\rho$ meson,
\begin{equation}
V_{\mu}=
\left(
  \begin{array}{cccc}
    \frac{1}{\sqrt{2}}(\rho^{0}+\omega) & \rho^{+}                             &  K^{*+}     & \bar{D}^{*0} \\
    \rho^{-}                            & \frac{1}{\sqrt{2}}(-\rho^{0}+\omega) &  K^{*0}     & D^{*-}       \\
     K^{*-}                             & \bar{K}^{*0}                         &  \phi       & D^{*-}_s     \\
     D^{*0}                             & D^{*+}                               &  D^{*+}_{s} & J/\psi       \\
  \end{array}
\right)_{\mu},
\end{equation}
and $B$ represents the 20-plet of the proton,
\begin{align}
&B^{121}=p,                        ~~~~~B^{122}=n,                                        ~~~~~ B^{132}=\frac{1}{\sqrt{2}}\Sigma^{0}-\frac{1}{\sqrt{6}}\Lambda,\nonumber\\
&B^{213}=\sqrt{\frac{2}{3}}\Lambda,~~~~~B^{231}=\frac{1}{\sqrt{2}}\Sigma^{0}+\frac{1}{\sqrt{6}}\Lambda,~~~~~B^{232}=\Sigma^{-},\nonumber\\
&B^{233}=\Xi^{-},B^{311}=\Sigma^{+},~~~~~B^{313}=\Xi^{0},~~~~~B^{141}=-\Sigma_c^{++},\nonumber\\
&B^{142}=\frac{1}{\sqrt{2}}\Sigma^{+}_{c}+\frac{1}{\sqrt{6}}\Lambda_c,~~~~~B^{143}=\frac{1}{\sqrt{2}}\Xi^{'+}_c-\frac{1}{\sqrt{6}}\Xi_c^{+},\nonumber\\
&B^{241}=\frac{1}{\sqrt{2}}\Sigma_c^{+}-\frac{1}{\sqrt{6}}\Lambda_c,~~~~~B^{242}=\Sigma_c^{0},\nonumber\\
&B^{243}=\frac{1}{\sqrt{2}}\Xi^{'0}_c+\frac{1}{\sqrt{6}}\Xi_c^{0},~~~~~B^{341}=\frac{1}{\sqrt{2}}\Xi^{,+}_c+\frac{1}{\sqrt{6}}\Xi_c^{+},\nonumber\\
&B^{124}=\sqrt{\frac{2}{3}}\Lambda_c,~~~~~B^{234}=\sqrt{\frac{2}{3}}\Xi_c^{0},~~~~~B^{314}=\sqrt{\frac{2}{3}}\Xi_c^{+},\nonumber\\
&B^{342}=\frac{1}{\sqrt{2}}\Xi^{'0}_c-\frac{1}{\sqrt{6}}\Xi_c^{0},~~~~~B^{343}=\Omega_c^{0},\nonumber\\
&B^{144}=\Xi^{++}_{cc},~~~~~B^{244}=-\Xi^{+}_{cc},~~~~~B^{344}=\Omega_{cc}\label{eqw2},
\end{align}
where the indices $i,j,k$ of $B^{ijk}$ denote the quark content of the baryon fields with the identification $1\leftrightarrow{u}, 2\leftrightarrow{d}, 3\leftrightarrow{}s$, and $4\leftrightarrow{}c$.   Following the strategy of Ref.~\cite{Montana:2017kjw}, the Lagrangian for the bottom hadrons can be easily obtained by replacing the charm baryons in Eq.~(\ref{eqw2}) with the bottom ones.

 To compute the $\rho{}KK$ and $\pi{}KK^{*}$ vertices, we also need the following effective Lagrangians
\begin{align}
&{\cal{L}}_{KK\rho}=ig_{KK\rho}[\bar{K}(\partial_{\mu}K)-(\partial_{\mu}\bar{K})K]\vec{\tau}\cdot\vec{\rho}^{\mu},\\
&{\cal{L}}_{K^{*}K\pi}=i\frac{g_{K^{*}K\pi}}{\sqrt{2}}K^{*\dagger}_{\mu}[(\vec{\tau}\cdot\vec{\pi})\partial^{\mu}K-\partial^{\mu}(\vec{\tau}\cdot\vec{\pi})K]\label{eqw1},
\end{align}
where
\begin{equation}
K=
\left(
  \begin{array}{cc}
     K^{+} \\
     K^{0}\\
    \end{array}
\right),~~~~~
\bar{K}=
\left(
  \begin{array}{cc}
     -K^{-} \\
     \bar{K}^{0}\\
    \end{array}
\right)
\end{equation}
with $\vec{\tau}$ being the usual Pauli matrices and $\vec{\rho}^{\mu}$ the field operators of the $\rho$ meson.
In the $SU(3)_f$ limit, the coupling constant satisfies $g_{KK\rho}=2.9$, which is determined by the
 Kawarabayashi-SuzukiRiazuddin-Fayyazuddin relation~\cite{Kawarabayashi:1966kd}.  The coupling $g_{K^{*}K\pi}$
 is fixed from the strong decay width of $K^{*}\to{}K\pi$.  With the help of Eq.~(\ref{eqw1}), the two-body decay width
 $\Gamma(K^{*+}\to{}K^{0}\pi^{+})$ is related to $g_{K^{*}K\pi}$ as
 \begin{align}
 \Gamma(K^{*+}\to{}K^{0}\pi^{+})=\frac{g^2_{K^{*}K\pi}}{6\pi{}m^2_{K^{*+}}}{\cal{P}}_{\pi{}K^{*}},
 \end{align}
where ${\cal{P}}_{\pi{}K^{*}}$ is the three-momentum of the $\pi$ in the rest frame of the  $K^{*}$.
Using the experimental strong decay width we obtain $g_{K^{*}K\pi}=4.61$~\cite{Tanabashi:2018oca}.

Putting all the pieces together, we obtain the following amplitudes,
\begin{align}
{\cal{M}}^{1/2}_{a}&(\Xi_b^{*-}\to\Lambda_b^0K^{-})=\sqrt{3}g_{\Xi_b^{*}\Sigma_b\bar{K}}gg_{KK\rho}\int\frac{d^4k_1}{(2\pi)^4i}\nonumber\\
                          &\times\Phi[(k_1\omega_{\bar{K}^{0}}-k_2\omega_{\Sigma_b^{-}})^2]\bar{u}(p_1)\gamma_{\mu}\frac{k\!\!\!/_1+m_{\Sigma_b^{-}}}{k_1^2-m^2_{\Sigma^{-}_b}}\Gamma{}u(k_0)\nonumber\\
                          &\times\frac{1}{k_2^2-m^2_{\bar{K}^0}}(k_2^{\nu}+p^{\nu}_{2})\frac{-g^{\mu\nu}+q^{\mu}q^{\nu}/m^2_{\rho^{-}}}{q^2-m^2_{\rho^{-}}},\\
{\cal{M}}^{1/2}_{b}&(\Xi_b^{*-}\to\Lambda_b^0K^{-})=\frac{\sqrt{3}}{2}g^2_{\Xi_b^{*}\Sigma_b\bar{K}}gg_{KK\rho}\int\frac{d^4k_1}{(2\pi)^4i}\nonumber\\
                         &\times{}\Phi[(k_1\omega_{K^{-}}-k_2\omega_{\Sigma_b^{0}})^2]\bar{u}(p_1)\gamma_{\mu}\frac{k\!\!\!/_1+m_{\Sigma_b^{0}}}{k_1^2-m^2_{\Sigma^{0}_b}}\Gamma{}u(k_0)\nonumber\\
                          &\times\frac{1}{k_2^2-m^2_{K^{-}}}(k_2^{\nu}+p^{\nu}_{2})\frac{-g^{\mu\nu}+q^{\mu}q^{\nu}/m^2_{\rho^{0}}}{q^2-m^2_{\rho^{0}}},\\
{\cal{M}}^{1/2}_{c}&(\Xi_b^{*-}\to\Xi_b^{0(')}\pi^{-})={\cal{K}}g^2_{\Xi_b^{*}\Sigma_b\bar{K}}gg_{K^{*}K\pi}\int\frac{d^4k_1}{(2\pi)^4i}\nonumber\\
                          &\times{}\Phi[(k_1\omega_{\bar{K}^{0}}-k_2\omega_{\Sigma_b^{-}})^2]\bar{u}(p_1)\gamma_{\mu}\frac{k\!\!\!/_1+m_{\Sigma_b^{-}}}{k_1^2-m^2_{\Sigma^{-}_b}}\Gamma{}u(k_0)\nonumber\\
                          &\times\frac{1}{k_2^2-m^2_{\bar{K}^0}}(k_2^{\nu}+p^{\nu}_{2})\frac{-g^{\mu\nu}+q^{\mu}q^{\nu}/m^2_{K^{*-}}}{q^2-m^2_{K^{*-}}},\\
{\cal{M}}^{1/2}_{d}&(\Xi_b^{*-}\to\Xi_b^{0(')}\pi^{-})=-{\cal{H}}g^2_{\Xi_b^{*}\Sigma_b\bar{K}}gg_{K^{*}K\pi}\int\frac{d^4k_1}{(2\pi)^4i}\nonumber\\
                          &\times{}\Phi[(k_1\omega_{{K}^{-}}-k_2\omega_{\Sigma_b^{0}})^2]\bar{u}(p_1)\gamma_{\mu}\frac{k\!\!\!/_1+m_{\Sigma_b^{0}}}{k_1^2-m^2_{\Sigma^{0}_b}}\Gamma{}u(k_0)\nonumber\\
                          &\times\frac{1}{k_2^2-m^2_{{K}^{-}}}(k_2^{\nu}+p^{\nu}_{2})\frac{-g^{\mu\nu}+q^{\mu}q^{\nu}/m^2_{K^{*0}}}{q^2-m^2_{K^{*0}}},
\end{align}
\begin{align}
{\cal{M}}^{3/2}_{a}&(\Xi_b^{*-}\to\Lambda_b^0K^{-})=-i\sqrt{3}g^2_{\Xi_b^{*}\Sigma_b\bar{K}}gg_{KK\rho}\int\frac{d^4k_1}{(2\pi)^4i}\nonumber\\
                          &\times{}\Phi[(k_1\omega_{\bar{K}^{0}}-k_2\omega_{\Sigma_b^{-}})^2]\bar{u}(p_1)\gamma_{\mu}\frac{k\!\!\!/_1+m_{\Sigma_b^{-}}}{k_1^2-m^2_{\Sigma^{-}_b}}\Gamma{}u^{\eta}(k_0)\nonumber\\
                          &\times\frac{1}{k_2^2-m^2_{\bar{K}^0}}(k_2^{\nu}+p^{\nu}_{2})\frac{-g^{\mu\nu}+q^{\mu}q^{\nu}/m^2_{\rho^{-}}}{q^2-m^2_{\rho^{-}}}k_1^{\eta},\\
{\cal{M}}^{3/2}_{b}&(\Xi_b^{*-}\to\Lambda_b^0K^{-})=i\frac{\sqrt{3}}{2}g^2_{\Xi_b^{*}\Sigma_b\bar{K}}gg_{KK\rho}\int\frac{d^4k_1}{(2\pi)^4i}\nonumber\\
                          &\times{}\Phi[(k_1\omega_{K^{-}}-k_2\omega_{\Sigma_b^{0}})^2]\bar{u}(p_1)\gamma_{\mu}\frac{k\!\!\!/_1+m_{\Sigma_b^{0}}}{k_1^2-m^2_{\Sigma^{0}_b}}\Gamma{}u^{\eta}(k_0)\nonumber\\
                          &\times\frac{1}{k_2^2-m^2_{K^{-}}}(k_2^{\nu}+p^{\nu}_{2})\frac{-g^{\mu\nu}+q^{\mu}q^{\nu}/m^2_{\rho^{0}}}{q^2-m^2_{\rho^{0}}}k_1^{\eta},\\
{\cal{M}}^{3/2}_{c}&(\Xi_b^{*-}\to\Xi_b^{0(')}\pi^{-})=i{\cal{K}}g^2_{\Xi_b^{*}\Sigma_b\bar{K}}gg_{K^{*}K\pi}\int\frac{d^4k_1}{(2\pi)^4i}\nonumber\\
                         &\times{}\Phi[(k_1\omega_{\bar{K}^{0}}-k_2\omega_{\Sigma_b^{-}})^2]\bar{u}(p_1)\gamma_{\mu}\frac{k\!\!\!/_1+m_{\Sigma_b^{-}}}{k_1^2-m^2_{\Sigma^{-}_b}}\Gamma{}u^{\eta}(k_0)\nonumber\\
                         &\times\frac{1}{k_2^2-m^2_{\bar{K}^0}}(k_2^{\nu}+p^{\nu}_{2})\frac{-g^{\mu\nu}+q^{\mu}q^{\nu}/m^2_{K^{*-}}}{q^2-m^2_{K^{*-}}}k_1^{\eta},\\
{\cal{M}}^{3/2}_{d}&(\Xi_b^{*-}\to\Xi_b^{0(')}\pi^{-})=-i{\cal{H}}g^2_{\Xi_b^{*}\Sigma_b\bar{K}}gg_{K^{*}K\pi}\int\frac{d^4k_1}{(2\pi)^4i}\nonumber\\
                          &\times{}\Phi[(k_1\omega_{{K}^{-}}-k_2\omega_{\Sigma_b^{0}})^2]\bar{u}(p_1)\gamma_{\mu}\frac{k\!\!\!/_1+m_{\Sigma_b^{0}}}{k_1^2-m^2_{\Sigma^{0}_b}}\Gamma{}u^{\eta}(k_0)\nonumber\\
                          &\times\frac{1}{k_2^2-m^2_{{K}^{-}}}(k_2^{\nu}+p^{\nu}_{2})\frac{-g^{\mu\nu}+q^{\mu}q^{\nu}/m^2_{K^{*0}}}{q^2-m^2_{K^{*0}}}k_1^{\eta},
\end{align}
where ${\cal{K}}=(\sqrt{3},\sqrt{2})$ and ${\cal{H}}=(\frac{\sqrt{6}}{2},1)$ for  amplitude ${\cal{M}}(\Xi_b^{*-}\to\Xi_b^{0}\pi^{-})$ and ${\cal{M}}(\Xi_b^{*-}\to\Xi_b^{'0}\pi^{-})$, respectively.

Once the amplitudes are determined, the corresponding partial decay widths can be obtained, which read,
\begin{align}
\Gamma(\Xi_b^{*}\to MB)=\frac{1}{2J+1}\frac{2}{8\pi}\frac{|\vec{p}_1|}{m^2_{\Xi_b^{*}}}\overline{|{\cal{M}}|^2},
\end{align}
where $J$ is the total angular momentum of the $\Xi_b^{*}$ state, the $|\vec{p}_1|$ is the three-momenta of the decay products in the center of mass frame,
the overline indicates the sum over the polarization vectors of the final hadrons, and $MB$ denotes the decay channel of $MB$, i.e.,$\Lambda_b\bar{K}$, $\Xi_b\pi$, $\Xi^{'}_b\pi$.

\section{RESULTS and Discussions}
In this work, we study the strong decays of the $\Xi_b^*$ to the two-body final states $\Lambda_b\bar{K}$, $\Xi_b\pi$,
and $\Xi_b^{'}\pi$ assuming that it is a $\bar{K}\Sigma_b$ molecular state.  In order to obtain the two body decay width
through the triangle diagrams shown in Fig.~\ref{fety}, we first compute the coupling constant $g_{\Xi_b^{*}\Sigma_b\bar{K}}$.
\begin{table}[h!]
\centering
\caption{ The coupling constant $g_{\Xi_b^{*}\Sigma_b\bar{K}}$ for different $J^P$ assignments with $\Lambda=0.9-1.1$ GeV.}\label{table0}
\begin{tabular}{cccc}
\hline\hline
   $J^P=\frac{1}{2}^{-}$     &~~  $J^P=\frac{1}{2}^{+}$    &~~  $J^P=\frac{3}{2}^{-}$(GeV$^{-1}$)   &~~ $J^P=\frac{3}{2}^{+}$(GeV$^{-1}$) \\ \hline
   $1.84-1.71$               &~~  $12.01-10.39$            &~~  $4.50-3.37$                         &~~ $3.87-2.95$                       \\ \hline
                     \hline
\end{tabular}
\end{table}

\begin{figure}[htbp]
\begin{center}
\includegraphics[bb=50 10 650 470, clip, scale=0.33]{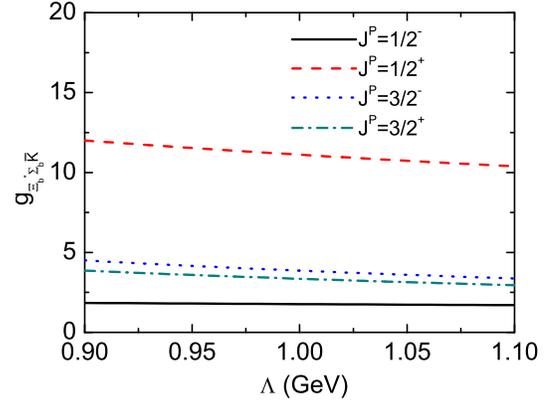}
\caption{The coupling constants of the $\Xi_b^{*}$ state with different $J^P$ assignments as a function of the parameter $\Lambda$.}\label{couplin-constant}
\end{center}
\end{figure}

With a value of the cutoff $\Lambda=0.9-1.1$ GeV, the corresponding coupling constants are listed in Table.~\ref{table0}.
The $\Lambda$ dependence of these coupling constants are  shown in Fig~\ref{couplin-constant}.  We note that they decrease
slowly with the increase of the cut-off, and the coupling constant is almost independent of  $\Lambda$ for the $J^P=1/2^{-}$
case, where the $\Xi_b^{*}$ is an $S-$wave $\bar{K}\Sigma_b$ molecular state.  It is consistent with the conclusion in Refs.~\cite{Dong:2010gu,Faessler:2007gv,Faessler:2007us,Dong:2008gb,Dong:2009uf,Dong:2009yp,Dong:2017rmg,Dong:2014ksa,Dong:2014zka,Dong:2013kta,Dong:2013iqa,Dong:2013rsa,Dong:2012hc,Dong:2011ys,Dong:2010xv,Dong:2009tg,Dong:2017gaw}  that for an $S-$wave loosely bound state the effective coupling strength of the bound state to its components
is insensitive to its inner structure.

\begin{figure}[htbp]
\begin{center}
\includegraphics[bb=40 30 700 500, clip,scale=0.37]{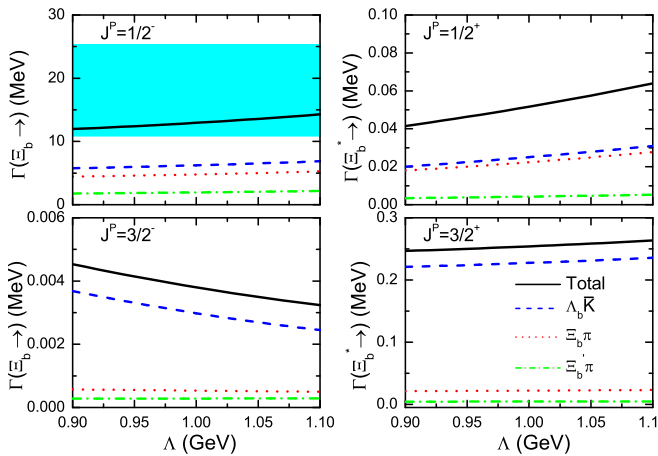}
\caption{Partial decay widths of the $\Xi_b^{*}\to{\Lambda_b\bar{K}}$ (blue dashed lines), $\Xi_b^{*}\to{\Xi_b\pi}$ (red dotted lines), $\Xi_b^{*}\to{\Xi^{'}_b\pi}$ (green dash-dotted lines), and the total decay width(black solid lines)
with different spin-parity assignments for the $\Xi_b^*$ as a function of the parameter $\Lambda$. The oycn
bands denote the experimental total width~\cite{Aaij:2018yqz}.}\label{width-ty}
\end{center}
\end{figure}

We show the dependence of the total decay width on the cutoff $\Lambda$ in Fig.~\ref{width-ty}.  In the present calculation,
we vary $\Lambda$ from 0.9 to 1.1 GeV.  In this $\Lambda$ range, the total decay width increases for the cases of
$J^{P}=1/2^{\pm}$ and $J^P=3/2^{+}$, while decreases for the $J^P=3/2^{-}$ case.  For the case of $J^P=1/2^{+}$, the predicted
total decay width is much smaller than the experimental total width, which disfavors such a spin-parity assignment for the
$\Xi_b^{*}$ in the $\bar{K}\Sigma_b$ molecular picture.  The $J^P=3/2^{+}$ assignment is not favored by our study as well. In this case
 the $\bar{K}\Sigma_b$ molecular state should be in $P-$wave.  Hence, the $J^P=3/2^{+}$ assumption for $\Xi_b^{*}$(6227)
can be excluded.   The $J^P=3/2^{-}$ case is disfavored due to the smallest width predicted, and a $D-$wave $\bar{K}\Sigma_b$
molecular with $J^P=3/2^{-}$ is difficult to form through long range meson exchanges.   In other words, the $J^P=1/2^{+}$ and $J^P=3/2^{\pm}$ $\Sigma_b\bar{K}$
molecular assumptions for the $\Xi_b^{*}$ are excluded.   Hence, only the assignment as an $S-$wave $\bar{K}\Sigma_b$ molecular state is possible for
the $\Xi_b^{*}$ based on the total decay width experimentally measured.

\begin{figure}[htbp]
\begin{center}
\includegraphics[scale=0.35]{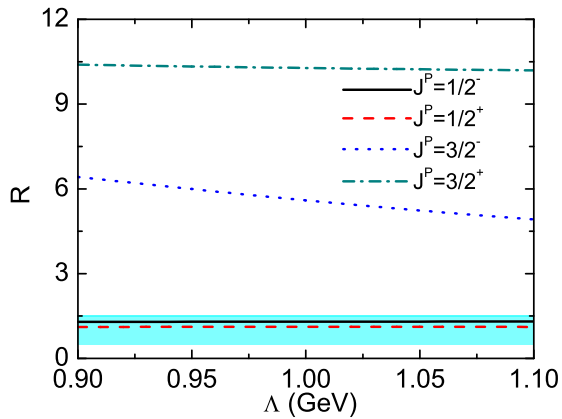}
\caption{Ratio of the partial decay widths B($\Xi_b^{*}\to{\Lambda_b\bar{K}}$)/B($\Xi_b^{*}\to{\Xi_b\pi}$) for different spin-parity assignments.
 The oycn band denotes the experimental result~\cite{Aaij:2018yqz}.}\label{width-ratio}
\end{center}
\end{figure}

In Fig.~\ref{width-ty}, we show the partial decay widths of the $\Xi_b^{*}\to\Lambda_b\bar{K}, \Xi_b\pi$, and $\Xi_b^{'}\pi$ as a function of the
cutoff parameter $\Lambda$.   The two-body decays in the $\Xi_b^{*} \to \bar{K}\Lambda_b,\Xi_b\pi$, and $\Xi^{'}_b\pi$ channels are insensitive to the
cut-off parameter $\Lambda$.   The transition $\Xi_b^{*} \to \bar{K}\Lambda_b$ and $\Xi_b^{*} \to\Xi_b\pi$ are the main decay channels in the $J^{p}=1/2^{\pm}$
case, the sum of which almost saturates the total width.  For the $J^P=3/2^{\pm}$ assignment, the transition $\Xi_b^{*}\to\Lambda_b\pi$ provides an dominant
contribution to the total decay width.  Since the phase space is small compared with the other two channels, the $\Xi_b^{'}\pi$ two-body decay width  is the smallest in
the four spin-parity assignments.

\begin{table*}[htbp!]
\centering
\caption{ Partial decay widths of $\Xi^{*}_{b}\to\Lambda_b\bar{K}$, $\Xi_b\pi$, $\Xi^{'}_b\pi$, and the total decay width $\Gamma_{total}$ with
$\Lambda=0.9-1.1$ MeV, in comparison with the results of  the  quark model~\cite{Wang:2018fjm,Chen:2018orb,Aliev:2018lcs} and the total width
obtained from the LHCb experiments~\cite{Aaij:2018yqz}.  The relative decay
ratio of the $\Lambda_b\bar{K}$ and $\Xi_b\pi$ channels are also shown in the eleventh row.  All masses and widths are in units of MeV.}\label{tablewidth}
\begin{tabular}{ccccccccccccccccccccc}
\hline\hline
             ~~&\multicolumn{1}{c}{This work}         &~~&\multicolumn{2}{c}{Reference\cite{Wang:2018fjm}}  &~~ &\multicolumn{2}{c}{Reference\cite{Chen:2018orb}}              &
&\multicolumn{1}{c}{Referrence\cite{Aliev:2018lcs}}      &  &\multicolumn{1}{c}{Exp.\cite{Aaij:2018yqz}} \\\cline{2-2} \cline{4-5} \cline{7-8} \cline{10-10} \cline{12-12}
             ~~&   $J^P=1/2^{-}$  &~~& $J^P=3/2^{-}$     &$J^P=5/2^{-}$                 &~~ &$J^P=3/2^{-}$ & $J^P=5/2^{-}$ &&$J^P=3/2^{-}$ \\
Decay models ~~&$\Xi_b^{*}$     &~~& $\Xi_b^{'}(6224)$ &$\Xi_b^{'}(6226)$             &~~ &~~ $\Xi^{'}_{b2}(6213)$~~~~~ &$\Xi^{'}_{b2}(6217)$&~~&$\Xi_b^{*}$                &   &$\Xi_b^{*}$\\\hline
$\Lambda_b\bar{K}$      &$5.76-6.87$   &  &5.9   &4.2  &&10.2&11.0   &&$1.58\pm0.45$\\
$\Xi_b\pi$              &$4.45-5.26$   &  &16.0  &16.4 &&11.4&11.7   &&$15.21\pm2.52$ \\
$\Xi^{'}_b\pi$          &$1.76-2.16$   &  &1.3   &0.6  &&1.0 &0.5    &&\\
$\Xi^{*}_b\pi$          &                    &  &      &     &&1.0 &1.7    &&\\
$\Xi^{'}_b(5945)\pi$    &                    &  &1.0   &3.2  &&&&&&\\
$\Xi_b(6096)\pi$        &                    &  &      &     &&$-$ &$-$    &&\\
$\Xi_b(6102)\pi$        &                    &  &      &     &&$-$ &$-$    &&\\
$Ratio$                 &$1.29-1.30$   &  &0.37  &0.26 &&0.89&$0.94$ &&0.10  && $0.5-1.5$\\
$Total$                 &$11.97-14.28$&  &24.2  &24.4 &     &23.6&24.9&&$16.79\pm2.57$ &  &$18.1\pm5.4\pm1.8$\\
\hline\hline
\end{tabular}
\end{table*}

 In Table~\ref{tablewidth}, we  list the decay widths of the $\Xi_b^{*}(6227)$ in the $J^P=1/2$ assignment with $\Lambda=0.9-1.1$ GeV.   The ratio of the partial decay widths in the
$\Lambda_b\bar{K}$ and $\Xi_b\pi$ channels are shown in the eleventh line.
For comparison, we show the results of the quark models~\cite{Wang:2018fjm,Chen:2018orb,Aliev:2018lcs} as well.
In Ref~\cite{Wang:2018fjm} and Ref~\cite{Aliev:2018lcs}, although the $\Xi_b^{*}$ state can decay into both $\bar{K}\Lambda_b$ and $\pi\Xi_b$
channels, the partial decay width of the $\bar{K}\Lambda_b$ mode is much smaller than that of the $\pi\Xi_b$ mode.  It contradicts with
the experimental data, where the ratio of the partial decay widths of the $\Xi_b^{*}$ into $\bar{K}\Lambda_b$ and
$\pi\Xi_b$ channels is about 0.5-1.5.   In Ref.~\cite{Chen:2018orb} the new $\Xi_b^{*}$ state can be well interpreted as a conventional three quark
state in comparison with the experimental total width and the ratio of the partial decay widths of the $\Xi_b^{*}$ into $\bar{K}\Lambda_b$
and $\pi\Xi_b$.  One should note that the difference between the results of Ref.~\cite{Chen:2018orb} and Refs.~\cite{Wang:2018fjm,Aliev:2018lcs} mainly originates from
the different choices on  fixing some parameters accounting for the strength of the quark-meson couplings and/or the  harmonic oscillator (SHO) wave functions.

 In Fig.~\ref{width-ratio}, we show the  ratio of the partial decay widths into
$\Lambda_b\bar{K}$ and $\Xi_b\pi$.  For the $J^P=1/2^{+}$ case, the ratio is about
1.06 but the total decay width is much smaller than the experimental value.  The ratio for the $J^P=3/2^{\pm}$ cases are
obviously inconsistent with the experimental results.  Therefore, our results, in terms of not only  the total decay width but also the ratio of the partial decay widths into
$\Lambda_b\bar{K}$ and $\Xi_b\pi$, support the $\Xi_b^{*}$ state as a $S-$wave $\bar{K}\Sigma_b$ molecular state.

\section{summary}
We studied the strong decays of the newly observed $\Xi_b^{*}$ baryon into $\bar{K}\Lambda_b^{0}, \Xi_b\pi$, and $\pi^{-}\Xi_b^{'}$  with different spin-parity assignments and assuming that it is a $\bar{K}\Sigma_b$ molecular structure.  With the couplings
constants between the $\Xi_b^{*}$ and its components determined by the composition condition,  we calculated the partial decay widths into
 $\bar{K}\Lambda_b^{0}$, $\Xi_b\pi$, and $\pi^{-}\Xi_b^{'}$ final states through triangle diagrams in an effective Lagrangian approach.
In such a picture, the decays $\Xi_b^{*}\to\bar{K}\Lambda_b^0$, $\Xi_b^{*}\to\pi\Xi_b$, and $\Xi_b^{*}\to\pi\Xi_b^{'}$ occur by exchanging
$\rho$ and $K^{*}$ mesons.   We found that both the total decay width and the ratio of the partial decay widths into
$\Lambda_b\bar{K}$ and $\Xi_b\pi$ can be reproduced with the assumption that the $\Xi_b^{*}$ is an $S-$wave $\bar{K}\Sigma_b$ bound
state with $J^P=1/2^{-}$, while the $P-$ and $D-$wave $\bar{K}\Sigma_b$ assignments are excluded. Our study shows that
the experimental measurement of the spin-parity of
the $\Xi_b^*$ will be able to tell whether it is a molecular state or a conventional three-quark state.

\section*{Acknowledgements}

YH thank Qi-Fang L\"{u} for valuable discussions. This work is partly supported by the National Natural
Science Foundation of China under Grants No.11522539, No.11675228, and No.11735003.


\end{document}